\documentclass[prc,floatfix,unsortedaddress,superscriptaddress,showpacs,showkeys]{revtex4}
\usepackage[dvips]{graphicx}
\usepackage[dvips,leftbars]{changebar}
\begin{document}

\title{FaCE: a tool for Three Body Faddeev calculations with core excitation }

\author{I.J. Thompson}
\affiliation{Department of Physics, University of Surrey, Guildford GU2 7XH, U.K.}

\author{F.M. Nunes}
\affiliation{
National Superconducting Cyclotron Laboratory, Michigan State University,
East Lansing, MI 48824, U.S.A.}
\email{nunes@nscl.msu.edu}

\author{B.V. Danilin}
\affiliation{RRC The Kurchatov Institute, Moscow, Russia}

\date{\today}

\begin{abstract}
FaCE is a self contained programme, with namelist input, that solves
the three body Faddeev equations. It enables the inclusion of excitation of
one of the three bodies, whilst the other two remain inert.
It is particularly useful for obtaining the binding
energies and bound state structure compositions of light exotic nuclei
treated as three-body systems,
given the three effective two body interactions.
A large variety of forms for these interactions may be defined,
and supersymmetric transformations of these potentials may be calculated
whenever two body states need to be removed due to Pauli blocking.
\end{abstract}


\maketitle

\parindent 0pt

{\Large PROGRAM SUMMARY}

\textit{Title of program: }FaCE (Faddeev with Core Excitation)

\textit{Computers:} The code is designed to run on any unix/linux workstation or PC.

\textit{Operating systems:} Linux or UNIX

\textit{Program language used:} Fortran-90

\textit{Numerical libraries used:} Source code for 6 routines from the NAG and
BLAS libraries is included to enable independent compilation.

\textit{Memory required to execute with typical data:} 9 Mbytes of RAM memory
and 12 MB of hard disk space.

\textit{No. of bits in a word: }32 or\textit{ }64

\textit{No. of lines in distributed program, including test data, outputs, etc.:}
13944

\textit{Distribution format:} compressed tar file


\textit{Nature of physical problem:} The program calculates eigenenergies and
eigenstates for the three body problem by solving the Faddeev equations.

\textit{Method of solution:} Given the two body effective potentials
it performs the supersymmetric transformation in case where there are
forbidden states to be removed. The three body wavefunction is
expanded in hyperspherical coordinates, the hyper-angular part is a series
of jacobi polynomials and the hyper-radial part is written in terms of a
laguerre basis. Within this basis the three body matrix elements
are  calculated and the full three body Hamiltonian matrix is completed.
The diagonalization process is performed after various reductions
(isospin, orthonormal and Feshbach) to determine the energies.
Finally the three body wavefunction is reconstructed and other
bound state observables are calculated.

\textit{Typical running time:} 6  sec on a 1.7 GHz
Intel P4-processor machine.
\parindent 10pt

\pagebreak

{\Large LONG WRITE-UP}

\section{Introduction}

Radioactive nuclear beams have allowed
the exploration of the nuclear driplines (proton rich and neutron rich),
and unveiled exotic phenomena. Many of the properties of light exotic nuclei
have been well described within few body formalisms:
one neutron halos such as $^{11}$Be and $^{19}$C have been described
with two body  (core+N) models \cite{be11,c19};
Borromean systems such as $^6$He and $^{11}$Li  have been well
modelled as three body (core+N+N) systems \cite{phyrep,li11};
and $^8$He has been successfully
accounted for within a five body picture \cite{he8}. Although in the early
days these models assumed all participants were inert, and concentrated on
treating the few body dynamics exactly, the advantage
of retaining degrees of freedom of the core was soon realised \cite{be12}.
Few body wavefunctions, solutions to the few body Hamiltonian with core
excitation, should contain the main components of any microscopic
calculation, with the advantage of its simplicity.

Few body models have become extremely useful in the field of light
radioactive nuclei, not only from the structure perspective, but
mainly for the purpose of reaction modelling \cite{reactions}.
Some important consequences were found when extracting radii from reaction cross sections \cite{jim},
when analysing transfer reactions for extracting spectroscopic factors \cite{ganil}
or when studying elastic and inelastic scattering \cite{crespo}.
Many of the features contained in the few body structure models
are essential for a good description of the reaction process.

In this paper, we present a self contained program that provides a
solution to a general three body problem where one of the clusters
is allowed to excite. In Section II a brief overview of
the construction of the three body basis is presented. In section III the
matrix elements required for the standard interaction are given.
Section IV discusses matrix reduction methods (useful for big calculations)
and then the diagonalisation procedure.
Section VI contains a summary of the observables that are
calculated in FaCE. In Section VII specific comments on the
program and the input manual is provided. Finally in Section VIII
we illustrate the use of FaCE with three physical examples.

\section{The three body basis}

Our intention was to develop a general tool to handle the bound state
properties of a nucleus well described as a three body system {\it i+j+k}
where one of the particles {\it is allowed to excite}.
FaCE is based on solving the Faddeev equations \cite{faddeev} with a
hyperspherical formulation of the general three body
problem \cite{gronwall,delves,phyrep}.
\begin{figure}[h]
\includegraphics*[width=15cm]{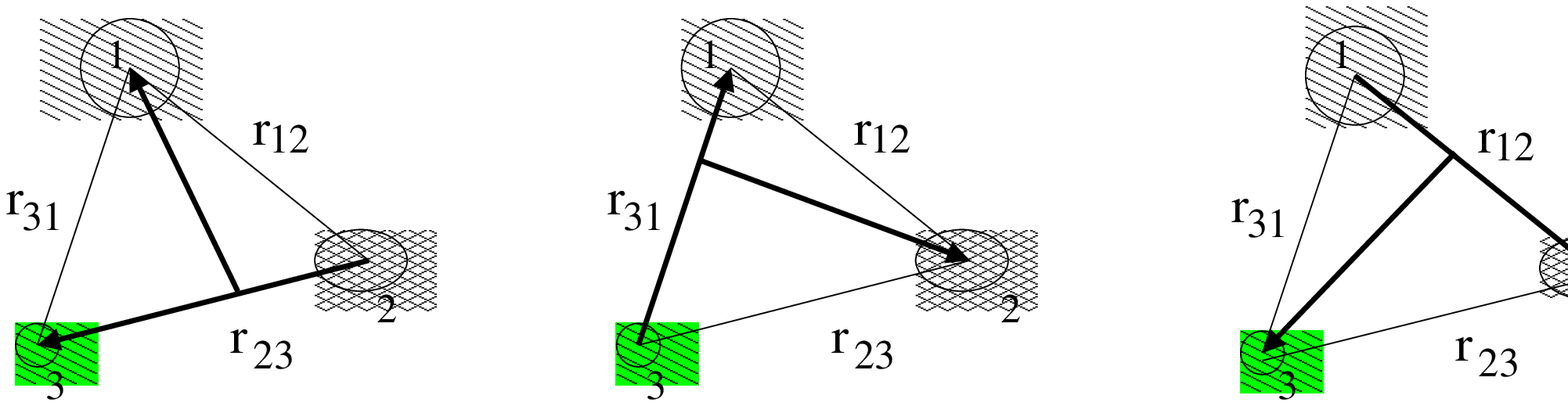}\\
\includegraphics*[width=15cm]{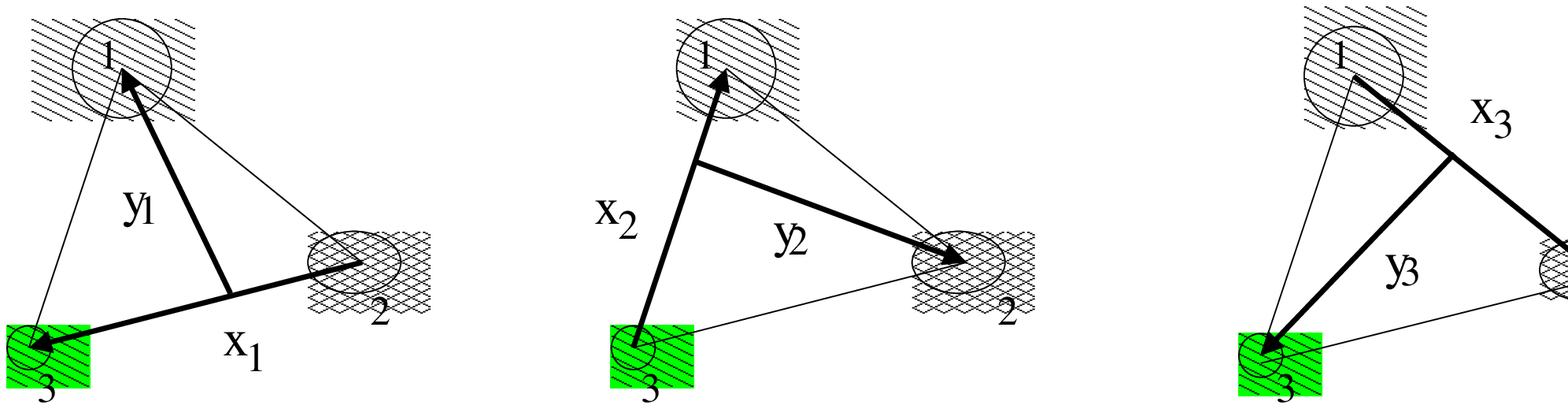}
\caption{Three sets of Jacobi Coordinates used in the Faddeev Formalism.}
\label{jac0}
\end{figure}

The Faddeev equations define three component wave functions $\Psi_i^{JM}$,
such that the full three-body wavefunction is
$\Psi^{JM} = \Psi_1^{JM}(x_1,y_1)+\Psi_2^{JM}(x_2,y_2)+\Psi_3^{JM}(x_3,y_3)$.
Here, the components $\Psi_i$ are functions of their own `natural' Jacobi coordinate
pairs $i$ (as in Fig. \ref{jac0}), and
are solutions of the Faddeev coupled equations:
\begin{eqnarray}
(T_1+h+V_1-E)\Psi_1^{JM}&=& - V_1(\Psi_2^{JM}+\Psi_3^{JM}) \nonumber \\
(T_2+h+V_2-E)\Psi_2^{JM} &=& - V_2(\Psi_3^{JM}+\Psi_1^{JM})   \label{eq:fad1} \\
(T_3+h+V_3-E)\Psi_3^{JM} &=& - V_3(\Psi_1^{JM}+\Psi_2^{JM}) \nonumber \;.
\end{eqnarray}
These equations contain $h=\sum_i h_i$, the sum of the intrinsic Hamiltonians of each particle $h_i$,
the relative kinetic energies in each coordinate set $T_i=T_{xi}+T_{yi}$ and the two body
interactions between the corresponding pair $V_i=V_{jk}(r_{jk})$
(both the Coulomb and nuclear interactions).
The indexes $i,j,k$ run through (1,2,3) in circular order.

The distances between each pair of particles
\cbstart{}%
$\vec r_{jk}$,  and the distance between the centre of mass of the pair
and the corresponding third particle (represented in Fig. (\ref{jac0})
by the thin lines),
\cbend{}%
can be  expressed in terms of the Jacobian coordinates
$(\vec x_i, \vec y_i)$ where $\vec{x}_i  =  \sqrt{A_{jk}}\; \vec{r}_{jk}$
 and $\vec{y}_i  =  \sqrt{A_{(jk)i}}\; \vec{r}_{(jk)i}$:
\cbstart{}%
\begin{equation}
\vec r_{jk} = \vec r_j - \vec r_k  \;\;
\mbox{~and~}
\vec{r}_{(jk)i} =  \vec r_i - (A_j\vec r_j +A_{k}\vec r_k)/(A_j+A_k)
\ .
\end{equation}
\cbend{}%
Note that the reduced masses are defined by $A_{jk} = \frac{A_j A_k}{A_j + A_k}$
and $A_{(jk)i} = \frac{(A_j+A_k) A_i}{A_i + A_j + A_k}$ with $i,j,k \in (1,2,3)$
where $A_i = \frac{m_i}{m}$ with  $m=1$ a.m.u. and $m_i$ the mass of particle
$i$ in a.m.u. In FaCE we will use  {\bf X} to refer to the pair ($x_1,y_1$),
{\bf Y} to refer to ($x_2,y_2$) and {\bf T} to refer to ($x_3,y_3$).

FaCE allows the user to include core excitation of one of the particles.
Let us assume one particle $c\in (1,2,3)$ has low lying excited states strongly
coupled to the ground state, and which are likely to have important roles
in the three body system. Particle {\em c} is then  treated as the core,
and its internal coordinates $\hat \xi_c$ must be added to the set of Jacobi
coordinates to define the full quantum state of the system.
The intrinsic Hamiltonian of the core determines a set of eigenstates $\phi_{s_c}$
and eigenvalues $\varepsilon_{s_c}$,
\begin{equation}
\hat h_{c}(\hat \xi_c) \; \phi_{s_c}(\hat \xi_c) = \varepsilon_{s_c} \;
\phi_{s_c}(\hat \xi_c) \;\;.
\end{equation}
The model then expands the total wavefunction of the system in terms
of these $\phi_{s_c}$ states, and factorizes the core degrees of freedom from the
Jacobi coordinates in each Faddeev component:
$$\Psi_i^{JM}(x_i,y_i,\hat \xi_c) = \sum_{s_c} \; \phi_{s_c}(\hat \xi_c)
\;\;\psi_{s_c}(x_i,y_i) \;,$$ with $i=1,2,3$.
Here $\psi_{s_c}$ contains the radial, angular and spin of the remaining two particles
relative to the chosen core. This model is advantageous if only a small number
of core states $\phi_{s_c}$ is required to describe the system accurately,
which is normally true  for systems close to the driplines.


FaCE uses the hyperspherical method to convert two-dimensional partial
differential equations into a set of coupled one-dimensional equations.
The Jacobi coordinates $(x_i,y_i)$ are transformed into the
hyperspherical coordinates (hyper-radius $\rho_i$ and hyper-angle $\theta_i$)
defined as
\begin{equation}
\rho^2  =  x_i^2 +y_i^2 = \sum_{i}^{3} A_i r_i^2
\hspace{0.5cm} \mbox{and}\hspace{0.5cm}
\theta_i  =  \arctan(\frac{x_i}{y_i}) \;\;.
\end{equation}
The {\it hyper-radius} is invariant under translations, rotations and
$(i,j)$ permutations, and is directly related to the overall size of the nucleus
whereas the {\it hyper-angle} contains radial correlations and
is related to the relative magnitude of the two Jacobi coordinates.
The hyper-radius is the same for all $i$={1,2,3}, this being
a basic advantage offered by the hyperspherical coordinate system,
while the hyper-angle $\theta_i=\arctan(\frac{x_i}{y_i})$ is different for
the various {\bf X, Y, T}  bases.

The transformation from a Jacobian coordinate set to
a hyperspherical coordinate system does not affect the angular
and spin variables, nor the degrees of freedom of the core.
Isospin dependence is not explicitly introduced since typically the
interactions to be used have a fixed isospin.

\begin{figure}[h]
\includegraphics*[width=15cm]{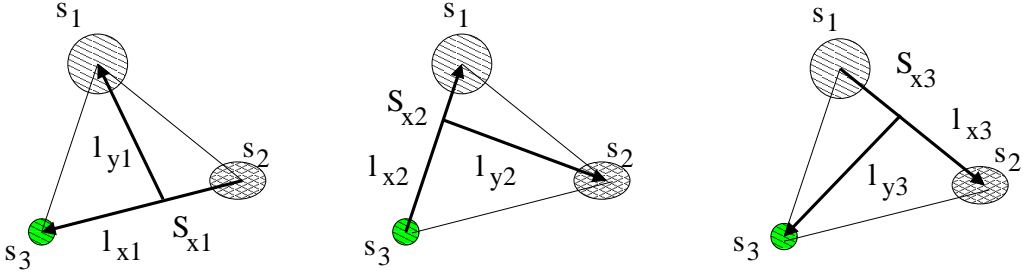}
\caption{Jacobi coordinates and the notation for the corresponding angular momenta.}
\label{jac2}
\end{figure}

For a given Jacobi set $(x_i,y_i)$, we need to define and couple together
the associated orbital angular momenta $(l_{xi},l_{yi})$,
as well as the states and the spins of the three particles
$s_i,s_j,s_k$, as shown in Fig. \ref{jac2}.
The core particle will have in FaCE an index for its different excited states,
but for the presentation below we assume that the spin value $s_c$ suffices to identify its state.
A partial wave decomposition for each Faddeev
component uses the following  the coupling order
\\
\begin{eqnarray}
\label{eq:partial}
 \Psi_{i}^{JM} & =& \sum_{l_{xi},l_{yi},L_i} \;\; \sum_{s_j,s_k,S_{xi}}
 \;\; \sum_{J_i,s_i}
\psi^{i,L_i S_{xi} s_i J}_{l_{xi} l_{yi}} (x_i,y_i)\;\;
     |i: \{  (l_{xi},l_{yi})L_i, \; (s_j,s_k)S_{xi}\} J_i;s_i \rangle^J\\
      & \equiv & \sum_{\alpha_i} \psi^{i,J}_{\alpha_i} (x_i,y_i) \;\; |i:~ \alpha_i \rangle^J
      \nonumber
\end{eqnarray}
with the abbreviation
$\alpha_i \equiv\{(l_{xi},l_{yi})L_1, \; (s_j,s_k)S_{xi}\} J_i;s_i$
for the quantum numbers of each component $i$. One of the $s_i,
s_j, s_k$ in $\alpha_i$ will identify the excitable core state $s_c$.

The two-dimension radial wavefunction  $\psi^{i,J}_{\alpha_i}
(x_i,y_i)$ is next expanded in the hyperspherical variables. The
separation between hyper-angle and hyper-radial dependence of the
wavefunction  makes use of the fact that the hyper-angle functions,
eigensolutions of the  hyper-angular (centrifugal) part of the
three-body kinetic energy operator \cite{delves}, are
explicitly defined in terms of the Jacobi polynomials:
\begin{eqnarray}
\psi^{i,J}_{\alpha_i} (x_i,y_i)& = & \rho_i^{-\frac{5}{2}}
\sum_{K_i} \; \chi^{i,J}_{\alpha_i K_i } (\rho) \;\;
\varphi_{K_i}^{l_{xi} l_{yi}}(\theta_i)\;\;,
\label{eq:hwf1}\\
\mbox{with} \;\;\;\;\;
\varphi_{K_i}^{l_{xi} l_{yi}}(\theta_i) & = & N_{K_i}^{l_{xi} l_{yi}} \;
(\sin\theta)^{l_{xi}}  \;  (\cos\theta)^{l_{yi}} \;\;
P_{n_i}^{l_{xi} + 1/2, l_{yi} + 1/2}(\cos2\theta_i) \;\;.
\label{eq:hwf2}
\end{eqnarray}
Here $ P_{ni}$ is the Jacobi polynomial, $N_{K_i}$ is a normalisation coefficient
and $K_i$ is the \mbox{{\it hyper-angular-momentum}}
directly related to the order of the corresponding
Jacobi polynomial $K_i=l_{xi}+l_{yi}+2n_i$ ($n_i$=0,1,2,...).
In order to simplify the notation we will omit whenever possible the total angular
momentum and projection labels $JM$ from the wavefunctions.

Introducing this expansion in the Faddeev Equations, and
performing the hyper-angular integration,  one obtains a set of
coupled equations for the wave functions
$\chi^i_{\alpha_iK_i}(\rho)$ of Eq. (\ref{eq:hwf1})
\begin{equation}
(T_\rho+L_{K_i}(\rho)-E)\chi^i_{\alpha_iK_i}(\rho) +
\sum_{j\alpha_jK_j}V^{ij}_{\alpha_iK_i,\alpha_jK_j}(\rho)
           \chi^j_{\alpha_jK_j}(\rho) = 0,
\label{coup-eq}
\end{equation}
where $T_\rho=-\frac{\hbar^2}{2m} \frac{d^2}{d\rho^2}$, and the
centrifugal potential is
$L_{K_i}(\rho)=\hbar^2 (K_i+3/2)(K_i+5/2)/(2m\rho^2)$.
The couplings are the hyper-angular integrations of the two-body interaction
$ V^{ij}_{\alpha_iK_i,\alpha_jK_j}(\rho)=
< \varphi_{K_j}^{l_{xj} l_{yj}}(\theta_j) | \hat V_{ij}|
\varphi_{K_i}^{l_{xi} l_{yi}}(\theta_i) > $.
In FaCE, these hyper-angular integrations are performed using
Gauss-Jacobi quadrature on a grid with $N_{jac}$ points
(defined in namelist {\bf grids} in the manual). Gauss-Jacobi
quadrature points are evenly spaced in hyper-angle.


In order to solve these coupled equations,
the hyper-radial behaviour is  expanded in terms of orthonormal basis functions
\begin{equation} \label{eq:rhobasis}
R_n(\rho)=\rho^{5/2}\rho_0^{-3}[n!/(n+5)!]^{1/2}L_n^5(z)\exp(-z/2)\ ,
\end{equation}
where $z=\rho/\rho_0$ with scaling radius $\rho_0$
and $L^\alpha_n(z)$ is an associated Laguerre polynomial,
as
\begin{equation}
\chi^{i,J}_{\alpha_i K_i } (\rho) =
\sum_{n=0}^{N_b} a^{in,J}_{K_i\alpha_i} R_n(\rho)\;.
\label{eq:hwf3}
\end{equation}
The potential matrix element integrals of the $R_n(\rho)$
functions are calculated using Gauss-Laguerre quadrature with
$N_{lag}$ points, which must be greater than the number
$N_b$ of basis polynomials. $N_b$ is set from the namelist
{\bf solve}, and $N_{lag}$ is set from the namelist
{\bf grids}, while the quadrature points and
weights are determined through finding numerically the roots of
$L^5_{N_{lag}}(z)=0$.

The kinetic energy matrix elements,
including the centrifugal barrier, are
$$
\langle R_n(\rho) | T_{K_i} | R_{n'}(\rho) \rangle =
\frac{\hbar^2}{2m} \left [
\frac{1}{2}  - \frac{\delta_{n{n'}}}{4} + \frac{n_<}{6} +
\frac{K_i(K_i+4)}{120} \left \{5(n_>-n_<+1)+n_>+n_<+1 \right \}
\right ]
$$
where $n_< = \min(n,{n'})$ and $n_> = \max(n,{n'})$.

After introducing the hyperspherical expansions Eqs. (\ref{eq:hwf1},\ref{eq:hwf3}) into
the Faddeev coupled equations Eq. (\ref{eq:fad1}), one arrives at a set of
simultaneous linear equations
\begin{equation} \label{eq:simeq}
   H {\bf a} = E {\bf a}
\end{equation}
for the coefficients  ${\bf a} \equiv \{a\}$.
We shall only be calculating
bound or pseudo-bound states, for which the wave functions $\chi(\rho)$
vanish at both $\rho=0$ and $\rho\rightarrow \infty$. This is guaranteed
by those same properties of the basis functions of Eq. (\ref{eq:rhobasis}).



\section{The three body matrix elements}

The complete wave function solution for a given $J$ (which will henceforth often be omitted)
is
\begin{eqnarray}
\label{eq:basis}
\Psi = \sum_{i=1}^3 \sum_{\alpha_i} \psi^{i}_{\alpha_i} (x_i,y_i) \;\; |i:~
\alpha_i \rangle\;,
\end{eqnarray}
where there is an implicit sum over hyper-moment K due to the
expansion of $\psi^{i}_{\alpha_i} (x_i,y_i)$ as in Eq. (\ref{eq:hwf1}).
The Hamiltonian matrix will therefore require overlap integrals of
the potentials between pairs of the overcomplete basis set $\{ |i:
\alpha_i\rangle \}$. We will need transformation matrices for the
rotations $|k: \alpha_k \rangle \rightarrow |i: \alpha_i \rangle$
clockwise, and $|i: \alpha_i \rangle \leftarrow |j: \alpha_j
\rangle$ anticlockwise, between the three Faddeev components.
Considering $(i,j,k)$ the circular order, the expressions that
allow the transformation (which conserves total angular moment
$L$ and hypermoment $K$)  between
Faddeev components in both directions are
\begin{eqnarray}
|i: \alpha_i \rangle & = & \sum_{S_t,S_{xk}} \sum_{L_k,l_{xk},l_{yk}} \sum_{J_k}
(-1)^{2(J-S_{xk}-s_k)+S_t+S_{xi}-s_i} \times
\hat{S_t}^2 \hat{S_{xk}} \hat{S_{xi}} \hat{J_k} \hat{J_i} \;
W(L_i,S_{xi},J,s_i;J_i,S_t) \nonumber \\
& \times & W(s_k,s_j,S_t,s_i; S_{xi}, S_{xk}) \; W(s_k,S_{xk},J,L_k;S_t,J_k) \;
RR(l_{xi}, l_{yi}; l_{xk}, l_{yk};L_k) \delta_{L_iL_k} |k: \alpha_k \rangle\;; \\
|i: \alpha_i \rangle & = & \sum_{S_t,S_{xj}} \sum_{L_j,l_{xj},l_{yj}} \sum_{J_j}
(-1)^{2(J-S_{xj}-s_j)+S_t-S_{xj}-s_j} \times
\hat{S_t}^2 \hat{S_{xj}} \hat{S_{xi}} \hat{J_j} \hat{J_i} \;
W(L_i,S_{xi},J,s_i;J_i,S_t) \nonumber \\
& \times & W(s_j,s_k,S_t,s_i; S_{xi}, S_{xj}) \; W(s_j,S_{xj},J,L_j;S_t,J_j) \;
RR(l_{xi}, l_{yi}; l_{xj}, l_{yj};L_j) \delta_{L_iL_j} |j: \alpha_j \rangle
\label{eq:norm}
\end{eqnarray}
where $RR(l_{xi}, l_{yi}; l_{xj}, l_{yj};L)$ are the
Raynal-Revai coefficients\cite{ray70}.
The two above equations introduce  two kinds of norm matrices  $N^{ij}$
and $\tilde{N}^{ik}$ such that $|i\rangle = N^{ij} |j\rangle$ and
$|i\rangle = \tilde{N}^{ik} |k\rangle$.
That is, the matrix elements of $N^{ij}$ are the basis-state
overlaps
$$
 N^{ij}_{\alpha_i\alpha_j} = \langle j: \alpha_j | i: \alpha_i \rangle \ ,
$$
so that
$N^{ij} N^{jk} = \tilde{N}^{ik}$.

There are several kinds of matrix elements needed for
the matrix $H$ in the Faddeev equations of Eq. (\ref{eq:simeq}).
The general potential we are considering is:
\begin{equation} \label{twobody-pot}
\hat V_{jk} = \hat V_i(x_i) = V^c_i(x_i) + \hat{SO} \; V^{so}_i(x_i) + \hat Q \; V^Q_i(x_i)
+\hat T \; V^T_i(x_i) + \hat{SS} \; V^{SS}_i(x_i) \;,
\end{equation}
where $V^c_i$ stands for the central interaction, $\hat{SO}$ and $V^{so}_k$
are the spin-orbit operator and the spin-orbit radial form-factor respectively,
$\hat Q$ and  $V^Q_i(x_i)$ are the tensor operator and radial shape
for the multipoles of the deformed potential, $\hat T$ and  $V^T_i(x_i)$
stand for the standard tensor operator \cite{brink} and radial dependence for the tensor NN interaction,
and finally $\hat{SS}$ and $V^{SS}_i$ are the spin-spin operator
and the corresponding form-factor. All these are included in FaCE.
The parameters for the corresponding radial form factors are defined
in namelist {\bf poten} (see manual for details).

The matrix elements of $\hat V_i(x_i)$ are preferentially
calculated between basis states of the same $i$ component.
The norm matrix elements $N^{ij}$ allow us to express general potential matrix elements in
mixed representations, in terms of the preferential Faddeev representation.
For example:
\begin{equation}
\langle  j: \alpha_j' \|  V_{i} \| i: \alpha_i \rangle =
\sum_{\alpha_i'}
N^{ij'}_{\alpha_i'\alpha_j'} \langle  i: \alpha_i' \|  V_{i} \
| i: \alpha_i \rangle\;,
\end{equation}
or, for cases when the potential is most easily presented
in one particular Jacobi coordinate set:
\begin{equation}
\langle  j':\alpha_j' \|  V_{i} \| j: \alpha_j \rangle = \sum_{\alpha_i'\alpha_i}
N^{i'j'}_{\alpha_i'\alpha_j'}
\langle  i:\alpha_i' \|  V_{i} \| i: \alpha_i \rangle N^{ji}_{\alpha_i\alpha_j}\;.
\end{equation}
In this section we first consider the angular and spin matrix elements,
and these will be later multiplied by numerical integrals over
the hyper-angle $\theta_i$ to obtain $V^{ij}_{\alpha_iK_i,\alpha_jK_j}(\rho)$
as in Eq. (\ref{coup-eq}) and
the hyper-radius $\rho$  after the expansion in Eq. (\ref{eq:hwf3}).
We will use $W_{ii'}$ to denote  the matrix elements over the angular momentum
basis states. We use:
\begin{equation}
\langle  i: {\alpha_i'} \|  V_{i} \| i: \alpha_i \rangle
 = W_{i i'}^c \; V^c_i(x_i) + W_{i i'}^{so} \; V^{so}_i(x_i) +
 W_{i i'}^Q \; V^Q_i(x_i) + W_{i i'}^T \; V^T_i(x_i) +
 W_{i i'}^{SS} V^{SS}_i(x_i) \;.
\end{equation}

The potential matrix element for the central part is diagonal
in all angular and spin variables. Next, let us consider the
spin-orbit part.
As all three particles may have spin, we have introduced the general
spin operator $\Sigma_i \equiv \Gamma_{ij} s_j + \Gamma_{ik} s_k$,
where $\Gamma_{ij}$ and $\Gamma_{ik}$ select which of the spins are to be
dynamically coupled, and with which relative strength.
The matrix elements for this operator are
\begin{eqnarray}
\hspace{-0.5cm} \langle  i\mbox{:} \alpha_i' \|  l_{xi} \cdot \Sigma_i \| i\mbox{:} \alpha_i \rangle & = &
\delta_{s_i's_i} \delta_{s_j's_j}\delta_{s_k's_k} \delta_{JJ'}
\delta_{J_i'J_i} \delta_{l_{yi}'l_{yi}} \delta_{l_{xi}'l_{xi}}
(-1)^{2J+3J_i'+l_{yi}+l_{xi}' + 2s_i + s_j + s_k} \nonumber \\
& \times & \hat{S_{xi}'} \hat{S_{xi}} \hat{L_{i}'} \hat{L_{i}} \hat{l_{xi}}
\sqrt{l_{xi} (l_{xi}+1)}
\left\{
\begin{array}{ccc} L_i'& l_i& 1 \\ S_{xi}& S_{xi}'& J_i \end{array} \right\}
\left\{
\begin{array}{ccc} L_i'& L_i& 1 \\ l_{xi}& l_{xi}'& l_{yi}' \end{array}\right\}
\nonumber \\
& \times & \left[ \Gamma_{ij} (-1)^{S_{xi}} \hat{s_{j}} \sqrt{s_{j} (s_{j}+1)}
\left\{ \begin{array}{ccc} S_{xi}'& S_{xi}& 1 \\ s_{j}& s_{j}'& s_k'
\end{array} \right\}
+
\Gamma_{ik} (-1)^{S_{xi}'} \hat{s_{k}} \sqrt{s_{k} (s_{k}+1)}
\left\{ \begin{array}{ccc} S_{xi}'& S_{xi}& 1 \\ s_{k}& s_{k}'& s_j'
\end{array} \right\}
\right] .
\label{eq:meso}
\end{eqnarray}

Typically the interaction between a deformed excitable nucleus
and another particle is expanded in multipoles. The essential
angular momentum operator is a tensor  interaction
of the type $C_Q(\hat l_{x}) \cdot C_Q(\hat s)$. Depending on the
Faddeev component and the subscript that specifies the deformed
nucleus, all forms of the operator
are needed: a) $C_Q(\hat l_{xi})\cdot C_Q(\hat s_i)$,
b) $C_Q(\hat l_{xi})\cdot C_Q(\hat s_j)$
and c) $C_Q(\hat l_{xi})\cdot C_Q(\hat s_k)$.
We next present the results for these matrix elements:
\\
\begin{eqnarray}
a) \;\;\; & &\langle  i: \alpha_i' \| C_Q(\hat l_{xi})
\cdot C_Q(\hat s_i) \| i: \alpha_i \rangle  =
 \delta_{s_j's_j}\delta_{s_k's_k} \delta_{JJ'}
\delta_{l_{yi}'l_{yi}} \delta_{S_{xi}'S_{xi}}
(-1)^{J'+2J_i+l_{yi}'+L_i'+L_i + s_i' + S_{xi}'}
\hat{J_{i}'} \hat{J_{i}} \hat{L_{i}'} \hat{L_{i}} \hat{l_{xi}'} \hat{l_{xi}}
\hat{s_{i}'} \nonumber \\
& & \times \left\{ \begin{array}{ccc} J_i'& J_i& Q \\ s_{i}& s_{i}'& J' \end{array} \right\}
\left\{ \begin{array}{ccc} J_i'& J_i& Q \\ L_{i}& L_{i}'& S_{xi}' \end{array} \right\}
\left\{ \begin{array}{ccc} L_i'& L_i& Q \\ l_{xi}& l_{xi}'& l_{yi}' \end{array} \right\}
\left( \begin{array}{ccc} l_{xi}' & Q & l_{xi} \\ 0 & 0 & 0 \end{array} \right)
\langle  s_{i'} \|  C_Q(\hat s_i) \| s_i \rangle\; ; \\
\nonumber \\
b) \;\;\; & & \langle  i: \alpha_i' \| C_Q(\hat l_{xi})
\cdot C_Q(\hat s_j) \| i: \alpha_i \rangle  =
 \delta_{s_i's_i}\delta_{s_k's_k} \delta_{JJ'} \delta_{J_i J_i'}
\delta_{l_{yi}'l_{yi}}
(-1)^{J'+l_{yi}' + S_{xi}'+S_{xi} + s_k' +s_j'}
\hat{L_{i}'} \hat{L_{i}} \hat{S_{xi}'} \hat{S_{xi}} \hat{l_{xi}'} \hat{l_{xi}}
\hat{s_{j}'} \nonumber \\
& & \times \left\{ \begin{array}{ccc} L_i'& L_i& Q \\ S_{xi}& S_{xi}'& J_{i}'
\end{array} \right\}
\left\{ \begin{array}{ccc} L_i'& L_i& Q \\ l_{xi}& l_{xi}'& l_{yi}'
\end{array} \right\}
\left\{ \begin{array}{ccc} S_{xi}'& S_{xi}& Q \\ s_j& s_j'& s_k
\end{array} \right\}
\left( \begin{array}{ccc} l_{xi}'& Q & l_{xi} \\ 0 & 0 & 0
\end{array} \right)
\langle  s_{j'} \|  C_Q(\hat s_j) \| s_j \rangle\; ; \\
\nonumber \\
c) \;\;\; & & \langle  i: \alpha_i' \| C_Q(\hat l_{xi})
\cdot C_Q(\hat s_k) \| i: \alpha_i \rangle  =
 \delta_{s_i's_i}\delta_{s_j's_j} \delta_{JJ'} \delta_{J_i J_i'}
\delta_{l_{yi}'l_{yi}}
(-1)^{J'+l_{yi}' + 2 S_{xi}' + s_k +s_j'}
\hat{L_{i}'} \hat{L_{i}} \hat{S_{xi}'} \hat{S_{xi}} \hat{l_{xi}'} \hat{l_{xi}}
\hat{s_{k}'} \nonumber \\
& & \times \left\{ \begin{array}{ccc} L_i'& L_i& Q \\ S_{xi}& S_{xi}'& J_{i}' \end{array} \right\}
\left\{ \begin{array}{ccc} L_i'& L_i& Q \\ l_{xi}& l_{xi}'& l_{yi}' \end{array} \right\}
\left\{ \begin{array}{ccc} S_{xi}'& S_{xi}& Q \\ s_k& s_k'& s_j' \end{array} \right\}
\left( \begin{array}{ccc} l_{xi}'& Q & l_{xi} \\ 0 & 0 & 0 \end{array} \right)
\langle  s_{k'} \|  C_Q(\hat s_k) \| s_k \rangle\; .
\label{eq:medef}
\end{eqnarray}
The matrix elements between different core states
$\langle  s_{i'} \|  C_Q(\hat s_i) \| s_i \rangle$ depends on the model
used. Under the assumption of a pure rotational model, these matrix elements
are given by:
\begin{equation}
\langle  s_{i'} \|  C_Q(\hat s_i) \| s_i \rangle = (-1)^{K-s_i'} \hat{s_{i}}
\left( \begin{array}{ccc} s_{i}'& Q & s_{i} \\ -K' &  0 & K \end{array} \right) \;,
\label{eq:merot}
\end{equation}
where $K$ is the quantum number for the rotational band. If the interaction
is $\ell-$dependent there is an ambiguity on the choice of the radial
form factor (which is defined at the multipole expansion of the deformed
potential). The parameter $lpot$ (see manual) controls this choice.

When considering the spin-spin interactions again there are
three possibilities depending on the Faddeev components:
a) $\langle  s_{i}' \| \vec s_i \cdot \vec s_j \| s_i \rangle$;
b) $\langle  s_{i}' \| \vec s_i \cdot \vec s_k \| s_i \rangle$ and
c) $\langle  s_{i}' \| \vec s_j \cdot \vec s_k \| s_i \rangle$.
After some algebra one can arrive at:
\begin{eqnarray}
a) \;\;\; \langle  i: \alpha_i' \|  \hat s_i \cdot \hat s_j \| i: \alpha_i \rangle &=&
\delta_{s_i's_i} \delta_{s_j's_j}\delta_{s_k's_k} \delta_{JJ'}
\delta_{L_i'L_i} (-1)^{3J+J_i'-J_i+L_{i}'+2S_{xi} - s_i + s_j + s_k}
\hat{J_{i}'} \hat{J_{i}} \hat{S_{xi}'} \hat{S_{xi}} \hat{s_{i}} \hat{s_{j}}
\nonumber \\
& \times & \sqrt{s_{i} (s_{i}+1)} \sqrt{s_{j} (s_{j}+1)}
\left\{ \begin{array}{ccc} J_i'& J_i& 1 \\ S_{xi}& S_{xi}'& L_i' \end{array} \right\}
\left\{ \begin{array}{ccc} S_{xi}'& S_{xi}& 1 \\ s_j& s_j' & s_k' \end{array} \right\}
\left\{ \begin{array}{ccc} s_i'& s_i& 1 \\ J_i & J_i' & J'  \end{array} \right\} \;;
\\
b) \;\;\; \langle  i: \alpha_i' \|  \hat s_i \cdot \hat s_k \| i: \alpha_i \rangle &=&
\delta_{s_i's_i} \delta_{s_j's_j}\delta_{s_k's_k} \delta_{JJ'}
\delta_{L_i'L_i} (-1)^{3J+J_i'-J_i+L_{i}'+S_{xi}+S_{xi}' - s_i' + s_j' + s_k}
\hat{J_{i}'} \hat{J_{i}} \hat{S_{xi}'} \hat{S_{xi}} \hat{s_{i}} \hat{s_{k}}
\nonumber \\
& \times &\sqrt{s_{i} (s_{i}+1)} \sqrt{s_{k} (s_{k}+1)}
\left\{ \begin{array}{ccc} J_i'& J_i& 1 \\ S_{xi}& S_{xi}'& L_i' \end{array} \right\}
\left\{ \begin{array}{ccc} S_{xi}'& S_{xi}& 1 \\ s_k & s_k' & s_j' \end{array} \right\}
\left\{ \begin{array}{ccc} s_i'& s_i& 1 \\ J_i & J_i' & J'  \end{array} \right\} \;;
\\
c) \;\;\; \langle  i: \alpha_i' \|  \hat s_j \cdot \hat s_k \| i: \alpha_i \rangle &=&
\delta_{s_i's_i} \delta_{s_j's_j}\delta_{s_k's_k} \delta_{JJ'}
\delta_{J_i'J_i} \delta_{S_{xi} S_{xi}'} \delta_{L_i'L_i}
(-1)^{S_{xi}+ s_j + s_k}
 \hat{s_{j}} \hat{s_{k}}
\nonumber \\
& \times & \sqrt{s_{j} (s_{j}+1)} \sqrt{s_{k} (s_{k}+1)}
\left\{ \begin{array}{ccc} s_j'& s_j& 1 \\ s_k & s_k' & S_{xi}'
\end{array} \right\} \;.
\label{eq:mess}
\end{eqnarray}

A realistic NN force contains a tensor interaction of the type
$T_2(s_js_k) \cdot C_2(l_{xi})$ which
also needs to be considered.  Below is the expression for these matrix elements
after working out the algebra:
\begin{eqnarray}
\langle  i: \alpha_i' \|  T_2(s_js_k) \cdot C_2(l_{xi} \| i: \alpha_i \rangle &=&
\delta_{s_i's_i} \delta_{s_j's_j}\delta_{s_k's_k} \delta_{JJ'}
\delta_{J_i'J_i} \delta_{l_{yi}'l_{yi}}
(-1)^{3J_i+l_{yi}-S_{xi}' } \hat 2 \hat{S_{xi}'} \hat{S_{xi}} \hat{L_{i}'}
\hat{L_{i}} \hat{l_{xi}}  \hat{l_{xi}'} \hat{s_{j}'} \hat{s_{k}'}
\sqrt{s_{j} (s_{j}+1)} \nonumber \\
& \times & \sqrt{s_{k} (s_{k}+1)}
\left\{ \begin{array}{ccc} S_{xi}'& S_{xi}& 2 \\ L_i & L_i' & J_{i}'
\end{array} \right\}
\left\{ \begin{array}{ccc} L_i' & L_i & 2 \\l_{xi}& l_{xi}'& l_{yi}'
\end{array} \right\}
\left( \begin{array}{ccc} l_{xi}'& 2 & l_{xi} \\ 0 & 0 & 0
\end{array} \right)
\left\{ \begin{array}{ccc} S_{xi}'& S_{xi}& 2 \\
s_j' & s_j & 1 \\s_k' & s_k & 1 \\  \end{array} \right\}.
\label{eq:metnn}
\end{eqnarray}

Few-body models often include effective three-body potentials to
describe the influence of dynamics not explicitly described by
two-body potentials. We have parameterised the simplest diagonal
form of such a potential:
\begin{eqnarray}
\langle  i': \alpha_i' \|  \hat{V}_3 \| i: \alpha_i \rangle =
\delta_{i'i} \delta_{\alpha_i'\alpha_i} V_3(\rho) \ .
\end{eqnarray}

\section{Pauli Blocking}

Often within a three-body calculation, it is necessary to
eliminate the Pauli forbidden  two-body bound states before
diagonalisation. This may be accomplished by several methods
\cite{dan98,thom00}: by projection operators inserted in
three-body Hamiltonian before diagonalisation, or by
transforming the two-body potentials in those partial wave channels
with deeply-bound forbidden states in a way that preserves
phase (spectral) equivalence.
We here adopt this second approach, and use supersymmetric
transformations of the two-body potentials in order to eliminate a
required set of bound states. All the parameters relative to the
method are specified in the namelist {\bf b2states} and the specific
characteristics of the two-body bound states to be calculated
are defined in {\bf b2state} (see manual for details).

Sometimes, it is useful to calculate the two body state generated
by a given effective interaction, or explore how to adjust
the two body interaction to obtain a given binding energy.
FaCE allows you to calculate bound states without feeding them
into the SUSY transformation subroutine (see {\bf b2states} in the manual
for details).

\subsection{Elimination of Two-body bound states}

A supersymmetric transformation  of the set of potentials
\cite{ss} enables the removal of an arbitrary bound state, while
keeping the spectral ($S$-matrix) equivalence of the initial and
the transformed Hamiltonians. In partition $k$, and in each
two-body spin-parity channel, the initial Hamiltonian $\hat{H}_0$
for the interaction of bodies $i,j$ couples $N$ two-body channels
for total angular momentum $j_k$. The two-body  equation is then
\begin{equation}
 \label{CC}\Bigl( - \frac{\hbar^2}{2\mu_{ij}} \Bigl[ \frac{d^2}{dr^2}
+ \frac{{ \ell_n}({ \ell_n} +1)}{r^2} \Bigr] - E \Bigr)
\phi_{n}(r) + \sum_{n'}^N V_{nn'}(r)  ~ \phi_{n'}(r) = 0 \
,\end{equation}
where $n$ is a channel index set $\{l_{x_k}, (s_i,s_j)S_{x_k};
j_k\}$, $r \equiv r_{ij}$, $\ell_n \equiv l_{x_k(n)}$, and
$\mu_{ij}$ is the reduced mass for bodies $i,j$. The threshold
energies $\epsilon_n = \varepsilon_{s_i} + \varepsilon_{s_j} $
are
included in the diagonal matrix elements $V_{nn}(r)$, in addition
to the couplings defined by Eq. (\ref{twobody-pot}). We start with
the real symmetric potential matrix $\hat{V}_0(r)=\{V_{n'n}(r)\}$
at each radius, and repeat the following supersymmetric
transformation for each bound state $p=1$ up to the number of
forbidden states $P$.

Let the column vector $\Phi_{p-1}^\lambda(r) =
\{\phi_{n}^\lambda(r)\}$ be the normalised ground state
eigensolution of  $H_{p-1}$ at real energy $E_\lambda$ below all
thresholds. By applying a double supersymmetric transformation to
$H_{p-1}$ we obtain a new Hamiltonian $H_p$ where the potential
matrix $\hat{V}_{p-1}(r)$ is replaced by $\hat{V}_p(r)$
\begin{equation} \label{susy}
\hat{V}_p(r)=
\hat{V}_{p-1}(r)-\frac{\hbar^2}{\mu_{ij}}\frac{d}{dr} \frac{\Phi_{p-1}^{\lambda
}(r)\Phi_{p-1}^{\lambda \dag}(r)} {\int_0^r \Phi_{p-1}^{\lambda \dag}(t)
\Phi_{p-1}^\lambda(t) dt}
\end{equation}

In the case of vanishing coupling between the channels near the
origin, 
it is possible to deduce the behaviour of the diagonal parts of
$\hat{V}_p(r)$ at small $r$. If the diagonal matrix of angular
moments $\{\ell_n\}$ has only one lowest element, say $\ell_1$,
such that $\ell_1<\ell_n$ for $n=2,3,...N$, in this channel (index
1) the additional term in the supersymmetric transformed potential
$\hat{V}_p(r)$  will have a singularity $\hbar^2
(2\ell_1+3)/(2\mu_{ij}r^2)$ at small $r$, which added to
the centrifugal term $\hbar^2 \ell_1(\ell_1 +1)/(2\mu_{ij}r^2)$
gives a new centrifugal term $\hbar^2
(\ell_1+2)(\ell_1+3)/(2\mu_{ij}r^2)$. The supersymmetric
transformation Eq. (\ref{susy}) in this case adds a repulsive core
at the origin, by increasing the orbital moment $\ell_1$ by 2
units. In all other channels the orbital moments $\ell_n$ are not
changed. Physically, this corresponds to the conservation of the
oscillator quanta $\Lambda=2n_r + \ell$ in the system: when
reducing the radial quantum number $n_r $ by one unit (removing
one level) we increase the orbital part $ \ell$ by two (for the
one channel case we satisfy the Levinson's theorem). If the
diagonal matrix of angular momenta $\{\ell_n\}$ has several lowest
equal elements, the increase of singularity is shared between
these channels, including their coupling potentials.

In FaCE, if supersymmetric transformations are used for any
partition $k$, then the transformations up to $\hat{V}_P(r)$
must be recalculated
for {\em all} desired two-body channels $j_k$ (all {\it jv}
in namelist {\bf b2state}).

\section{Solving the Faddeev equations}

The Faddeev equations Eq. (\ref{eq:fad1}) are solved, after expanding on the
hyper-angular  Eq. (\ref{eq:hwf2}) and hyper-radial Eq. (\ref{eq:rhobasis}) basis
functions, to find square-integrable solutions Eq. (\ref{eq:basis}) for
eigenenergies $E$ and eigenvectors ${\bf a}$. For $E<0$ these are bound states, whereas
for $E>0$ the eigen-solutions are `quasi-bound' states that form a
discrete representation of the continuum. These quasi-bound wave
functions may be used, for example in \cite{crespo},
in the calculations of breakup as inelastic excitations.

The physical normalisation of the wave functions
Eq. (\ref{eq:basis}) is $\langle \Psi | \Psi \rangle = 1$.
If $N$ is the whole normalisation matrix
$N^{ij}_{\alpha_i\alpha_j}$, the eigenvectors ${\bf a}$
are physically normalised when ${\bf a}^T N {\bf a} = 1$.
Note that some eigenvectors will be found that are non-physical,
having ${\bf a}^T N {\bf a} = 0$; in these there is a
cancellation between different Faddeev components, and they must be
omitted in all bound or breakup state analyses.

\subsection{Hamiltonian Reduction procedures}

The complete set Eq. (\ref{eq:fad1}) of Faddeev equations may be
reduced in a number of circumstances. FaCE has the option of `isospin' and
`orthonormal' reductions, which exactly reproduce a physically
chosen subset of the eigen-solutions, and also `Feshbach'
reduction, which is a method for approximating the effects of
high $K$ partial waves on the solutions. The choice of the reduction
method is made through {\it eqn} in the input namelist {\bf solve}
(see manual).

\subsubsection{Isospin Reduction}

Suppose bodies $j$ and $k$ are fermions which are isospin
states $T_z$ of some particle of isospin $T_{jk}$,
with $s_j = s_k$ half-integral. The requirement of
antisymmetrisation under exchange of these bodies is easily
satisfied if the partial wave set $\alpha_i$ {\em only} includes
those quantum number sets for which $l_{xi} + S_{xi} + T_{jk}$ is
{\em odd}. In this way, wave function components that are
symmetric under interchange of $j$ and $k$ are eliminated from the
basis set for this Faddeev component.

Furthermore, the remaining Faddeev components $\Psi_j$ and $\Psi_k$
are isospin mirrors of each other. Just
one of these wave functions needs to be included explicitly in the
equation set to be found numerically, since
\begin{equation}
  \Psi_j = -(-1)^{T_{jk}} P_{jk} \Psi_k
\end{equation} \label{eq:antisym}
where $P_{jk}$ is the operator permuting the coordinates of
particles $j$ and $k$.

The coupled equations
 \begin{equation}
 {\left( \begin{array}{ccc}
 T_i+h+V_i-E & V_i & V_i  \\
 V_j &  T_j + h+V_j-E & V_j \\
 V_k & V_k & T_k + h+V_k-E
 \end{array} \right)}
 {\left( \begin{array}{c} \Psi_i \\ \Psi_j \\ \Psi_k \end{array} \right)}
 =0
 \label{eq:matrix}
 \end{equation}
are now reduced to
 \begin{equation}
 {\left( \begin{array}{ccc}
 T_i+h+V_i-E  & V_i + V_i P_{jk} \\
 V_k &  T_k + h+V_k + V_kP_{jk}-E
 \end{array} \right)}
 {\left( \begin{array}{c} \Psi_i  \\ \Psi_k \end{array} \right)}
 =0 \ .
 \label{eq:matrix-reduced}
 \end{equation}
The permutation matrix elements are
$P_{jk}=(-1)^{(l_{xj}+S_{xj}-s_{2j}-s_{3j})}$.

\subsubsection{Orthonormal Reduction}

Since basis states $|i:~ \alpha_i \rangle$
in Eq. (\ref{eq:basis}) form an over-complete set,
the same set of physical eigen-solutions may be found by
transforming the basis set into an orthonormal one.

A rotation matrix $C$ may be found, for example by Gramm-Schmidt
orthonormalisation, such that $C^T N C = I$, so that the columns of $C$ are
vectors that are physically orthonormalised by the norm matrix $N$.
This same rotation
may be used to transform the Hamiltonian matrix of
Eq. (\ref{eq:simeq}). Defining $D = C^{-1} = C^T N$, then
$$
   D H C {\bf a} \equiv  H' {\bf a}' = E {\bf a}'
$$
is an orthogonal transformation of the original eigenvalue
problem, with the same eigenvalues. The original eigensolutions
may be regained as ${\bf a}' = C {\bf a}$.
The new matrix $H'$ is real and symmetric; this proves that the
eigenvalues of Eq. (\ref{eq:simeq}) are real even though $H$ is
not symmetric.

\subsubsection{Feshbach Reduction}
Another reduction method, also called the semi-adiabatic reduction method,
constructs  an effective coupling matrix at each $\rho$ value
using Feshbach's expression \cite{feshbach} for effective
interactions in a subspace.

Consider the set of $N$ coupled equations for the wave functions
$\chi^i_{\alpha_iK_i}$ as in Eq. (\ref{coup-eq}).
Take the subset of the equations of this system with largest $K_i$ and
core excitation energy $\varepsilon_{s_c}$. In these channels, an adiabatic
condition might be fulfilled, where the hyper-radial kinetic energy $T_\rho$ is small and can be
neglected. Thus  we have an option of keeping this kinetic
energy term in only the subset of channels $i=1,\cdots,M$, and of neglecting $T_\rho$
for $i=M+1,\cdots,N$. For each $\rho$ value, let us rewrite the system
Eq. (\ref{coup-eq}) in the following matrix form:
\begin{eqnarray}
\left (
\begin{array}{cccccccccc}
{\bf A} -  E & {\bf B}\\
{\bf C}      & {\bf D}-E
\end{array}  \right )
\left(
\begin{array}{c}
{\bf \chi^{(a)}}\\
{\bf \chi^{(b)}}
\end{array}
\right )
=0,
\label{matrix-eq}
\end{eqnarray}
where ${\bf A}$ contains the exact $T_\rho+L_{K_i}(\rho)$ terms,
but ${\bf D}$ contains only the $L_{K_i}(\rho)$ terms.
The ${\bf B}$ and ${\bf C}$ are the block off-diagonal matrices.
The solution vectors are ${\bf\chi^{(a)}} = (\chi_1 \cdots \chi_M)$
and ${\bf\chi^{(b)}} = (\chi_{M+1} \cdots \chi_N)$.

Solving the matrix Eq. (\ref{matrix-eq}) formally we obtain:
\begin{equation}
\chi^{(b)} = ({\bf D}-E)^{-1}{\bf C}\chi^{(a)},\label{chib}
\end{equation}
and substituting  Eq. (\ref{chib}) into our system Eq. (\ref{matrix-eq})
we get a reduced subset of coupled equations for $\chi^{(a)}$
\begin{equation}
\left({\bf A}-E + {\bf B}({\bf D}-E)^{-1}{\bf C}\right)\chi^{(a)}=0 \label{red-sub}.
\end{equation}

>From Eq. (\ref{red-sub}), we see that the reduction of the
coupled equations from $N\times N$ to a smaller $M\times M$ set consists in
adding a `Feshbach' term ${\bf B}({\bf D}-E)^{-1}{\bf C}$ to the
effective interaction in the retained subspace.

Strictly speaking, the Feshbach term should be recalculated for
every eigen-energy $E$, but in practice we calculate the Feshbach term once for the fixed
`Feshbach energy' $E=E_F$, which should be chosen near the
eigen-energy of the state of most interest,
such as the ground state energy. Variables {\it efesh} and {\it kmaxf}
in the input namelist {\bf solve} are the Feshbach energy and the
K-value above which the Feshbach approximation is introduced (see manual).

\subsection{Diagonalisation procedure}
The subroutine {\sc fadco} in FaCE evaluates
all the potential matrix elements as functions of $\rho$. After
the above possible reductions, the Hamiltonian matrix appearing in
the Faddeev Equations Eq. (\ref{eq:simeq}) is determined by
hyper-radial integrals using the radial basis function Eq.
(\ref{eq:rhobasis}).

For the general eigenvalue solution of Eq. (\ref{eq:simeq}), where $H$ is a
real matrix not necessarily symmetric, we use the subroutine
F02AGF from the Nag library, which proceeds via reduction to
Hessenberg form, to find all eigensolutions.

If only selected eigenvalues are required, and the input parameter {\it
meigs} (introduced in {\bf solve}) is non-zero and less than
the dimension of the $H$ matrix of Eq. (\ref{eq:simeq}),
a more efficient method is that of inverse iteration.
Starting from some energy $E_0$, and some initial guess ${\bf
a}^{(0)}$ for an eigenvector, the solution of the simultaneous
linear equations $ (H - E_0) {\bf a}^{(n)} = {\bf a}^{(n-1)}$ for
$n\ge 1$ will converge to the eigensolution with energy nearest
$E_0$. This method is effective for $E_0$ less than the ground
state energy, when there are no nearly-degenerate eigenvalues. It
may be generalised to finding the {\em several} eigenvalues
nearest $E_0$ by orthogonalising ${\bf a}^{(n)}$ at each iteration
to the set of eigenvectors already found.


\section{Computer programme and Input manual}

Given the description covered in the previous sections,
the FaCE manual is presented as a sequence of namelists with
explanatory names for the variables.
Nevertheless it is useful to remind the user that
the parameters delimiting the three body space are:
the number of Laguerre polynomials $N_b$ for the
hyper-radial part, together with the Jacobi polynomials for
the hyper-angular part $N_{jac}$; the maximum angular momentum that
are to be taken into account in each Faddeev partition
$l_{xmax}(i),l_{ymax}(i)$, and the number of $K$-harmonics $K_{max}(l,i)$.

The source code is distributed with separate makefiles for Sun
f90 compilers (standing alone, or with system Sun Performance Library and Nag
libraries) and for Linux, where there are Intel ifort and Portland Group
pgf90 makefiles, the latter optionally with system Lapack and Blas
libraries. The suitable one of these makefiles should be renamed to
`makefile'.

FaCE uses F02AGF, M01DAF and M01ZAF routines from the
Nag library, and DGETRF, DGETRS and DGEMM from the Blas library.
The original source codes for the Nag and Blas subroutines
are contained in the package for compilation where not otherwise
available.

\subsection{Input namelists}

\begin{itemize}

\item
{\bf {\&}fname}\\
{\it nfile[A*20], desc[A*80]}\\
{\it nfile} is the root name of the output files  and {\it desc} is a heading
that should describe and identify the run.

\item
{\bf {\&}scale}\\
{\it amn, hc [r*8]}\\
{\it amn} is the mass of the nucleon and {\it hc} is the planck constant multiplied
by the velocity of light in [MeV fm].

\item
{\bf {\&}nuclei}\\
{\it name(1:3)[A*8], mass(1:3), z(1:3), radius(1:3)[r*8]}\\
Each of the three interacting nuclei are characterised by their
name, mass, charge and radius.

\item
{\bf {\&}identical}\\
{\it id(1:3)[logical], iso(1:3)[r*8]}\\
If {\it id(j)} is true then the interacting pair in partition {\it j} are
\cbstart{}%
2 identical particles. This variable affects the choice of the basis:
if particles are identical, an isospin reduction is performed.
{\it iso} contains the isospin of the interacting pair to be used when
$id(j)$ is true, to omit non-antisymmetric partial waves in that partition.
\cbend{}%

\item
{\bf {\&}total}\\
{\it ngt[int], gtot(1:ngt)[r*8], gparity(1:ngt)[int]}\\
{\it ngt} $\rightarrow$ the number of states to be calculated,
{\it gtot} and {\it gparity} hold their total spin and parity(even$=+1$, odd$=-1$) respectively.

\item
{\bf {\&}particles}\\
{\it ns(1:3)[int], spin(1:ns,1:3)[r*8], parity(1:ns,1:3)[int],
energy(1:ns,1:3)[r*8]}\\
This namelist defines intrinsic properties of the three bodies.
For each of the 3 nuclei: {\it ns(j)} specifies the number of
states to be included for nucleus {\it j}. For each of its {\it ns(j)} states,
one should specify the spin, parity(+1/-1) and its energy relative
to the energy of the ground state of that nucleus {\it j}.

\item
{\bf {\&}em}\\
{\it corek(1:3), def(2:mmultipoles,1:3)
 [r*8]}

This namelist contains the electromagnetic information on each of the
three bodies.  \\
{\it corek(j)} $\rightarrow$ projection of the spin of nucleus $j$ in its rest frame \\
{\it def(q,j)} $\rightarrow$ deformation length in the $q$ multipole of nucleus $j$.

\item
{\bf {\&}waves}\\
This namelist contains the definition of the channels per partition.

{\it sym(3)[A*1]} the name (e.g 'T', 'X' or 'Y') of each partition.

{\it auto(3)[logical]}
$\rightarrow$ if `T' then FACE generates automatically
the channels allowed given maximum quantum numbers for each Faddeev partition.
Otherwise, quantum numbers for $nc(j)$ channels are explicitly read in.

If {\it auto} is false: \\
{\it nc(3)[int]}
$\rightarrow$ number of channels per partition (less than $mfchan$).\\
{\it lx(mfchan,3)[int], ly(mfchan,3)[int], lt(mfchan,3)[int],
sx(mfchan,3)[r*8], jp(mfchan,3)[r*8]}
$\rightarrow$ determine the quantum numbers associated with all channels for each partition
in the following coupling order $|[(l_x,l_y)l_t, (s_i,s_j)s_x] J_p, icy \rangle$  \\
{\it icy(mfchan,3)[int],icx1(mfchan,3)[int],icx2(mfchan,3)[int]}
$\rightarrow$ the state of the spectator, the first and the second interacting particle
for the given channel in each partition. The spin of a given state
of each of the bodies was defined in {\bf particles}. \\
{\it np(mfchan,3)[int]} $\rightarrow$ the number of $K$-harmonics.

If {\it auto} is true: \\
{\it lxmax(1:3)[int], lymax(1:3)[int], ltmax(1:3)[int], sxmax(1:3)[r*8], jabmax(1:3)[r*8], \\
kmaxa(1:3)[int], kmax(0:maxl,1:3)[int]} $\rightarrow$ these establish the maximum quantum
numbers allowed in each partition.
$kmaxa$ is the $K$ limit for all partial waves, and may be overridden by particular
$kmax$ specified.

\item
{\bf {\&}poten}\\
This namelist defines the radial behaviour of the potentials for each interacting
pair.

{\it detail(3)[A*80]} $\rightarrow$ information on the interaction between the interacting pair in that partition \\
{\it typc(3)[A*3], pa(6,3),ps(6,3),pp(6,3),pd(6,3),pf(6,3)[r*8]} $\rightarrow$ central interaction\\
{\it typso(3)[A*3],pso(6,3),psop(6,3),psod(6,3),psof(6,3)[r*8]} $\rightarrow$ spin-orbit interaction\\
{\it typss(3)[A*3],pss(6,3),psss(6,3),pssp(6,3),pssd(6,3),pssf(6,3)} $\rightarrow$ spin-spin
interaction\\
{\it typt(6,3)[A*3], pt(6,3)[r*8]}   $\rightarrow$ tensor interaction\\
{\it rcoul(3)[r*8],acoul(3)[r*8]} $\rightarrow$ the Coulomb interaction\\
{\it lpot(3)[int]} $\rightarrow$ is useful for $l-$dependent interactions
where there is an ambiguity on the radial form factor that should be
used for off diagonal couplings. If $lpot=0$, the radial form factor
corresponding to  the minimum $l_i,l_f$ is used;
$lpot=1$, the average is
taken; $lpot=2$ the maximum is used; and lpot=3 the final $l_f$
(corresponding to the left hand side of the matrix element) is used.
Finally, when  $lpot \ge 10$, the radial form factor for
off-diagonal coupling is determined by $l=lpot-10$,
throughout the whole calculation, leaving the monopole terms untouched.
\\

The form factor for the potentials between the interacting pair in each partition
is specified by type ({\tt gau}, {\tt ws}, {\tt rnp}, {\tt rnn}, {\tt nul})
and the potential parameters for  each partial wave
({\bf s},{\bf p},{\bf d},{\bf f} and  {\bf a} or no extra letter for all).
If the type is {\tt gau} then the interaction
is the sum of 3 Gaussians:
\begin{equation} \nonumber
V_{gau}^i(r) = \sum_{k=1,3,5} pa(k,i) \exp
   \left [ - \left ( \frac{r}{pa(k+1,i)}\right)^2 \right ]
\end{equation}
If type is {\tt ws} then the interaction is the sum of 2 Woods-Saxon:
\begin{equation} \nonumber
V_{ws}^i(r) =  \sum_{k=1,4} pa(k,i)
    \left [ 1 + \exp \left( \frac{r-pa(k+1,i)}{pa(k+2,i)}\right)\right ]^{-1}
\end{equation}
For the spin-orbit interaction, if type is 'ws' then the form factor is given
by the derivative of two  Woods-Saxon:
\cbstart{}%
\begin{equation} \nonumber
V_{ws}^i(r) =  \sum_{k=1,4}  \frac{ pso(k,i)}{r \; pso(k+2,i)}
\frac{ \exp( \frac{r-pso(k+1,i)}{pso(k+2,i)} )}
     {[ 1 + \exp(\frac{r-pso(k+1,i)}{pso(k+2,i)} )]^2}
\end{equation}
\cbend{}%

The Coulomb interaction for the interacting pair in partition (i) is that of a uniform
sphere with radius {\it rcoul(i)} and diffuseness {\it acoul(i)},
screened at radius {\it rscreen} with a Fermi function of diffuseness {\it ascreen}.

The operator for the tensor force is $12 \hat S_{12}$ as defined by Brink and
Satchler \cite{brink}.
The operator for the spin-spin force is the dot product of the spins of the
interacting pair $ \vec s_j \cdot \vec s_k $. The operator for the spin-orbit
is defined in {\bf gamso}. If the deformation of one of the interacting
particles in non zero then higher order multipoles will be automatically
added to the monopole interaction based on a spherical harmonic decomposition
of a deformed field.

\item
{\bf {\&}pot3b}\\
{\it typ3b,s3b(ngt),r3b(ngt),a3b(ngt),gtvary} $\rightarrow$ specifies the parameters for
the diagonal 3-body potential $V_3(\rho)$ if {\it typ3b} $\ne$ {\tt nul}.
If {\it gtvary}, then $ijt=1$ below, otherwise $ijt$ is the $J^\pi$ index 1 ... $ngt$.
If $typ3b = {\tt gau}$, then
\begin{equation} \nonumber
V_{3}(\rho) = s3b(ijt) \exp \left [ -\left( \frac{\rho}{r3b(ijt)} \right)^2 \right ]
\end{equation}
If $typ3b = {\tt ws}$, then
\begin{equation} \nonumber
V_{3}(\rho) = s3b(ijt) \left [ 1 + \exp\left( \frac{\rho-r3b(ijt)}{a3b(ijt)}  \right)\right ]^{-1}
\end{equation}

\item
{\bf {\&}gamso}\\
{\it gamso1, gamso2} $\rightarrow$ the spin-orbit matrix elements are calculated
using the following operator $ \Gamma_1 \vec l_x \cdot \vec s_1 +
\Gamma_2 \vec l_x \cdot \vec s_2 $
for each partition\\

\item
{\bf {\&}grids}\\
{\it rr [r*8], nlag,njac [int]
}
\\
This section contains radii for the expansions used.\\
{\it rr} $\rightarrow$ scaling parameter $\rho_0$ for the Laguerre basis \\
{\it nlag} $\rightarrow$ number of Laguerre quadrature points for the $\rho$ coordinate. \\
{\it njac} $\rightarrow$ number of Jacobi polynomials \\

\item
{\bf {\&}trace}\\
{\it
pripot,vadia [logical]}\\
Printing options:
$pripot$ prints the potential matrix elements,
$vadia$ prints the diagonalised coupling eigenvalues (energy surfaces).
All of these are printed in the output file with extension {\tt lis}.

\item
{\bf {\&}b2states}\\
{\it n2states[int],  dx,xmax,
[r*8], ipc,lmax,nk [int], rnode,de [r*8]}, \\
Find {\it n2states} two-body states in the two-body potentials.
Use radial grid 0 to {\it xmax} with steps {\it dx}.
The {\it ipc, lmax, nk} and {\it rnode, de} are default values for each {\bf b2state}
namelist below.

\item
{\bf {\&}b2state} (repeated {\it n2states} times)\\
{\it pair,kind [int], de [r*8] 
ipc [int], test [logical], n,nvchan,l,lmax [int],s,jv,rnode [r*8],
search,rescale [logical], eigen,potential, fermi [r*8],
nomit [int],omit\_l{1:nomit} [int] ,omit\_s{1:nomit},omit\_j{1:nomit} [r*8],
omit\_c1{1:nomit},omit\_c2{1:nomit} [int]},\\
Find two-body eigenstate in the potential {\it pair}, of {\it kind}=`occup'
to be used for pauli blocking via the susy transformation;
{\it kind}=`transfer' if one needs to check the properties of a particular
two body state, or
{\it kind}=`pot' if one needs to calculate numerically the potential
for a particular two body partial wave set, without excluding it.
This last option is needed, because whenever there are any occupied states,
the potentials for all partial waves sets in that partition
need to be calculated numerically.
\\
{\it ipc} = trace level, {\it test}=T to ignore this state after finding it.\\
Wave functions will have {\it n} nodes in channel $\ell$={\it l} up to radius {\it
rnode}, from a set of $\ell\le${\it lmax} using coupling order
$| \ell, (s_1s_2)s; jv\rangle$. The eigenenergy is {\it eigen} in monopole potential
multiplied by {\it potential}, where {\it search}=`E' or `V' to search for energy or
\cbstart{}%
potential factor respectively.
Energies {\it eigen} are negative for bound states. Only bound
states can be found when search ='E'.
\cbend{}%
Use {\it nomit}$>$0 to specifically omit some partial waves from the coupled channels
set.  \\
{\it fermi} $<$0, to  exclude bound states up to that valence
energy,\\
{\it fermi} $>$0, to  exclude the nint({\it fermi}) number of lowest-energy bound states.\\

\item
{\bf {\&}solve}\\
{\it eig, eimin,eimax,efesh [r*8], eqn[A*1], cfiles[logical],
nbmax, meigs(1:ngt),kmaxf
[int]}\\
This namelist is related with the type of equation to be solved: \\
{\it eig}, {\it eimin} and {\it eimax} are the target, minimum and maximum eigenenergies
to search for states, \\
{\it eqn} asks for  the reduction of the full equation:
eqn= 'F' stands for Faddeev,
eqn={\it sym(i)} (defined in {\bf waves}) performs an orthonormal transformation
to the $i$ basis,  \\
{\it nbmax} = number of functions in the radial expansion, must be $\le$ {\it nlag}, \\
{\it meigs} is the number of eigenstates to  calculate ({\it meigs}=0 is to find all
eigenstates).\\
{\it kmaxf} $\ge$ 0 for Feshbach reduction of coupled equations
at each hyper-radius to $K\le$ {\it kmaxf}, using eigenenergy estimate {\it efesh}.\\
{\it cfiles} = `T', to write {\tt mel} and {\tt spec} files to be fed into
an independent program of the coupled equations
(e.g the program sturmxx \cite{sturmxx}).\\

\end{itemize}

\subsection{Outputs}

\begin{itemize}

\item{\bf standard output}\\
The standard output contains the information about the three nuclei,
the partial waves to be included,  the two body potentials, the parameters
used in the expansions. If the run uses supersymmetric potentials,
the details regarding the two body bound states to be excluded
are printed out. Next, FaCE prints the angular momentum information
about all the possible channels, the Gauss-Laguerre grid, the details about
the reductions performed and the corresponding new reduced
set of channels. Finally the energy, the radii, and the probabilities
associated with each channel are  given for each calculated state:
values for L-summed probabilities, and summed probabilities for
each core-state are also included.
As JJ coupling is easier to compare with
the shell model basis, FaCE performed the LS-JJ
transformation and prints out the probabilities of the main JJ
components at the very end of the file.

The following files are produced with {\em filename} = {\em nfile}
in the {\bf fname} namelist.
\item{\bf filename.wf}\\
This file contains the hyper-radial wavefunction for the states $J^\pi$
calculated. It first contains the channels that are included in this output
(very small components are left out) followed by the
wavefunction in format $r,wf(i,r)$. This file can be easily plotted.

\item{\bf filename.nl}\\
FaCE rewrites into filename.nl the input as is read.

\item{\bf filename.lis}\\
This file contains extensive information on the various steps
of the calculation. It contains the two body potentials for each partition,
the algebra matrix elements presented in section III, the various radial
potential couplings for the various hyper-radii belonging to the
Gauss-Laguerre grid, the normalization and permutation matrices,
and the probability of the various configurations in long format.
\end{itemize}

\section{Examples of calculations}

\subsection{$^{12}$Be}

Input file {\em be12gptdefk4.in} and a shortened version of the
output file {\em be12gptdefk4.out} are provided below. The full files
are included in the electronic file distribution.
This example models $^{12}$Be as a three-body cluster of two neutrons
outside a $^{10}$Be core. The core is deformed and allowed to excited to
its first $2^+$ state. This example is similar to that in \cite{be12}
although here we use shallow core-n potentials, to most simply avoid
Pauli-forbidden two-body states.

\subsubsection{Example: be12gptdefk4.in}
\footnotesize
\begin{verbatim}
 &fname   nfile='be12gptdefk4' desc= 'be12gptdefk4: n+n+be10 using gptnn and be10-n , Kmax=4' /
 &scale amn=939. hc=197.3/
 &nuclei name=  'n','n','10be' mass= 1 1 10    z= 0 0 4    radius=0 0 2.30  /
 &identical id=F, F, F,  iso=0.5, 0.5, 0. /
 &total  ngt = 1, gtot(1)=0.0, gparity(1)=+1 /
 &particles
  ns(1)=1, spin(1,1)= 0.5, parity(1,1)=1, energy(1,1)=0.0,
  ns(2)=1, spin(1,2)= 0.5, parity(1,2)=1, energy(1,2)=0.0,
  ns(3)=2, spin(1,3)= 0.0, parity(1,3)=1, energy(1,3)=0.0,
           spin(2,3)= 2.0, parity(2,3)=1, energy(2,3)=3.368/
 &em
  corek(1)=0.5  def(2,1)=0.0 Qmom(1)=0.0 Mmom(1)=0.0
  corek(2)=0.5  def(2,2)=0.0 Qmom(2)=0.0 Mmom(2)=0.0
  corek(3)=0.0  def(2,3)=1.6638 Qmom(3)=0.0 Mmom(3)=0.0         def(4,3)=0 /

 &waves    auto(1)=T, kmaxa(1)=4,  lxmax(1)=2,
           auto(2)=T, kmaxa(2)=4, lxmax(2)=2,
           auto(3)=T, kmaxa(3)=4, lxmax(3)=2/
 &poten
   detail(1) ='n+10be'    typc(1) ='ws'
         ps(1,1)=-10.14 ps(2,1)=2.736, ps(3,1)=0.67
         pp(1,1)=-24.24 pp(2,1)=2.736, pp(3,1)=0.67
         pd(1,1)=-10.14 pd(2,1)=2.736, pd(3,1)=0.67
         lpot(1)=0
   typso(1)='ws'
         psop(1,1)=+25.72 psop(2,1)=2.736, psop(3,1)=0.67
         psod(1,1)=-25.72 psod(2,1)=2.736, psod(3,1)=0.67
   typss(1)='nul' typt(1) ='nul'

   detail(2) ='10be+n'    typc(2) ='ws'
         ps(1,2)=-10.14 ps(2,2)=2.736, ps(3,2)=0.67
         pp(1,2)=-24.24 pp(2,2)=2.736, pp(3,2)=0.67
         pd(1,2)=-10.14 pd(2,2)=2.736, pd(3,2)=0.67
         lpot(2)=0
   typso(2)='ws'
         psop(1,2)=+25.72 psop(2,2)=2.736, psop(3,2)=0.67
         psod(1,2)=-25.72 psod(2,2)=2.736, psod(3,2)=0.67
   typss(2)='nul' typt(2) ='nul'

   detail(3) = 'gptnn'    typc(3) ='gau'
         ps(1,3)=  560.0 ps(2,3)=0.8109 ps(3,3)=-390.7
         ps(4,3)=1.031 ps(5,3)=-1.501 ps(6,3)=3.205
         pp(1,3)=  9.335 pp(2,3)=  1.184  pp(3,3)=  -1.37
         pp(4,3)=  2.099 pp(5,3)=  0.1663 pp(6,3)=  3.193
         pd(1,3)=  560.0 pd(2,3)= 0.8109 pd(3,3)= -390.7
        pd(4,3)= 1.031 pd(5,3)= -1.501 pd(6,3)= 3.205
   typso(3)='gau'
         pso(1,3)= -114.5 pso(2,3)=0.9296
   typss(3)='nul'    typt(3) ='gau'
         pt(1,3)= 12.24 pt(2,3)= 1.539 pt(3,3)= -31.64
         pt(4,3)= 0.4039 pt(5,3)= 0.8111 pt(6,3)= 3.015
   /
 &pot3b  typ3b='nul', s3b=0., r3b=3.9    /
 &gamso  gamso1=1,0,1, gamso2=0,1,1               /

 &grids  rr=0.3 nlag=20 njac=40     /  Methods:
 &trace      pripot='T'   /
 &b2states n2states=0  dx=0.002 xmax=15.    /
 &solve   eimin = -5.0, eimax=3, eqn='F' nbmax=10 meigs=1 /
\end{verbatim}

\normalsize
\subsubsection{Output: be12gptdefk4.out}
\footnotesize
\begin{verbatim}
 FACE: version 0.12e

 Case file: be12gptdefk4

 be12gptdefk4: n+n+be10 using gptnn and be10-n , Kmax=4

  with constants:
  unit mass =   939.00 MeV;  hc =  197.300 MeV.fm =>  h2sm = 20.7281

           1       2       3
  Nuclei:  n       n       10be
  masses   1.0000  1.0000 10.0000
  charges     0.0     0.0     4.0
  radii    0.0000  0.0000  2.3000

  Identical 23: F,  31: F,  12: F,
    isospin   0.5     0.5     0.0

  1 coupled channels J,pi sets:  0.0+ #  1


  Particle 1: n        has 1 states:
   -- spin,parity,energy = 0.5+ @  0.0000 MeV
   -- Intrinsic K = 0.5  Quadrupole moment =   0.000 Magnetic moment =   0.000

  Particle 2: n        has 1 states:
   -- spin,parity,energy = 0.5+ @  0.0000 MeV
   -- Intrinsic K = 0.5  Quadrupole moment =   0.000 Magnetic moment =   0.000

  Particle 3: 10be     has 2 states:
   -- spin,parity,energy = 0.0+ @  0.0000 MeV 2.0+ @  3.3680 MeV
   -- Intrinsic K = 0.0  Quadrupole moment =   0.000 Magnetic moment =   0.000
   -- deformation lengths =   1.66380

  Partial waves:
  Component 1 X: lx,ly,lt <=  2 10 10, sx,jp <= 2.510.0, Kmax(all,0:lx) =  4 -1 -1 -1
  Component 2 Y: lx,ly,lt <=  2 10 10, sx,jp <= 2.510.0, Kmax(all,0:lx) =  4 -1 -1 -1
  Component 3 T: lx,ly,lt <=  2 10 10, sx,jp <= 1.010.0, Kmax(all,0:lx) =  4 -1 -1 -1


 POTENTIALS

 Potential 1 between n        and 10be    :
   n+10be
  Central potential of type 'ws ' ,
   for   s-waves: -10.14000   2.73600   0.67000   0.00000   0.00000   0.00000
   for   p-waves: -24.24000   2.73600   0.67000   0.00000   0.00000   0.00000
   for   d-waves: -10.14000   2.73600   0.67000   0.00000   0.00000   0.00000
   using lpot =   0
  Spin-orbit potential of type 'ws ' ,
   for   p-waves:  25.72000   2.73600   0.67000   0.00000   0.00000   0.00000
   for   d-waves: -25.72000   2.73600   0.67000   0.00000   0.00000   0.00000
    acting on n        with factor  1.0000, on 10be     with factor  0.0000
  Spin-spin  potential of type 'nul' ,
  Tensor     potential of type 'nul' ,

 Potential 2 between 10be     and n       :
   10be+n
  Central potential of type 'ws ' ,
   for   s-waves: -10.14000   2.73600   0.67000   0.00000   0.00000   0.00000
   for   p-waves: -24.24000   2.73600   0.67000   0.00000   0.00000   0.00000
   for   d-waves: -10.14000   2.73600   0.67000   0.00000   0.00000   0.00000
   using lpot =   0
  Spin-orbit potential of type 'ws ' ,
   for   p-waves:  25.72000   2.73600   0.67000   0.00000   0.00000   0.00000
   for   d-waves: -25.72000   2.73600   0.67000   0.00000   0.00000   0.00000
    acting on 10be     with factor  0.0000, on n        with factor  1.0000
  Spin-spin  potential of type 'nul' ,
  Tensor     potential of type 'nul' ,

 Potential 3 between n        and n       :
   gptnn
  Central potential of type 'gau' ,
   for   s-waves: 560.00000   0.81090-390.70000   1.03100  -1.50100   3.20500
   for   p-waves:   9.33500   1.18400  -1.37000   2.09900   0.16630   3.19300
   for   d-waves: 560.00000   0.81090-390.70000   1.03100  -1.50100   3.20500
   using lpot =   0
  Spin-orbit potential of type 'gau' ,
   for all-waves:-114.50000   0.92960   0.00000   0.00000   0.00000   0.00000
    acting on n        with factor  1.0000, on n        with factor  1.0000
  Spin-spin  potential of type 'nul' ,
  Tensor     potential of type 'gau' ,
      12.24000     1.53900   -31.64000     0.40390     0.81110     3.01500

 Three-body potential of type 'nul' ,


  METHOD parameters:

  Hyperradial parameters =  0.3000   20
  Hyperangular points =           40
  nbmax =           10 , eig =   -5.000000

  nlag =           20  recommend: >> nbmax =           10
  njac =           40  recommend: >> kmax/2 =            2
    (but much more, for repulsive-core interactions)

  Equations to solve = F=F  (F=Faddeev, T,X,Y-Schrodinger+orthoNormal)

  For  0.0+ search for    1 eigenstates near  -5.000 MeV,
  examine those between  -5.000 &   3.000 MeV

 ********************************************
 * Coupled channels set  1 for J,pi =  0.0+ *
 ********************************************


  Faddeev channel numbers required:           30           30           30

 X:
 ig 1   1   2   3   4   5   6   7   8   9  10  11  12  13  14  15  16
  i 1   1   2   3   4   5   6   7   8   9  10  11  12  13  14  15  16
  K 1   0   2   2   2   4   4   4   4   4   2   2   2   2   2   2   2
  L 1   0   0   1   0   0   1   0   1   0   2   1   2   2   2   2   2
 sx 1 0.5 0.5 0.5 0.5 0.5 0.5 0.5 0.5 0.5 1.5 1.5 1.5 1.5 2.5 2.5 2.5
 lx 1   0   1   1   0   2   2   1   1   0   0   1   1   2   0   1   2
 ly 1   0   1   1   0   2   2   1   1   0   2   1   1   0   2   1   0
 jp 1 0.5 0.5 0.5 0.5 0.5 0.5 0.5 0.5 0.5 0.5 0.5 0.5 0.5 0.5 0.5 0.5
 iz 1   1   1   1   1   1   1   1   1   1   2   2   2   2   2   2   2

<< .... 54 lines deleted .... >>

 All Gauss-Laguerre points:
 radii:        0.252     0.500     0.813     1.194     1.646     2.170     2.771
               3.452     4.217     5.073     6.026     7.086     8.263     9.573
              11.036    12.680    14.549    16.712    19.307    22.679

 Gauss-Laguerre points for nbmax =           10 :
 radii:        0.451     0.901     1.478     2.196     3.071     4.127     5.401
               6.952     8.896    11.527

   Calculating HH up to n =            2 , l1,l2 =            2            3
   Calculating matrix elements
    do potentials for partition             1
    do potentials for partition             2
    do potentials for partition             3
    allocate ww section of             1  Mb
    allocate ww files of 2 *            2  Mb
  For basis             1 , dnn =    1.048809
  For basis             2 , dnn =    1.048809
  For basis             3 , dnn =    1.414214
  eqn,id(:)=F  F  F  F
    allocate wm array of             8  Mb
  Matrix elements found in Gauss-Laguerre Basis
  Diagonalise matrix of size           990
  Search for             1  eigenstates near    -5.000000

 Inverse iteration to find   1 eigenstates nearest  -5.00000
 iteration    1 gives   -4.70616, change = -4.71E+00
 iteration    2 gives   -4.87763, change = -1.71E-01
 iteration    3 gives   -4.87759, change =  3.51E-05
 iteration    4 gives   -4.87759, change =  6.10E-07

 Found   0.0+ Evals at:     -4.87759

 Bound  0.0+ eig    1 =   -4.877592
  rms <rho> =  4.514 fm., rms matter radius =  2.471 fm.

  Probability norms in eigenstate:

   i:  K  L  sx lx ly jp  i#        normalised      permuted
  X:
    1  0  0 0.5  0  0 0.5 1 1 1    0.06145327    0.58052013
    2  2  0 0.5  1  1 0.5 1 1 1    0.04085356    0.29732071
    3  2  1 0.5  1  1 0.5 1 1 1    0.01151844    0.04324449
    4  2  0 0.5  0  0 0.5 1 1 1    0.00336356    0.00247767
    5  4  0 0.5  2  2 0.5 1 1 1    0.00004349    0.00563919
<< .... 25 lines deleted .... >>

   P(S) = 0.583931,  P(S') = 0.303015, P(P) = 0.045894, P(D) = 0.067025

  Y:
   31  0  0 0.5  0  0 0.5 1 1 1    0.06145327    0.58052013
   32  2  0 0.5  1  1 0.5 1 1 1    0.04085356    0.29732071
   33  2  1 0.5  1  1 0.5 1 1 1    0.01151844    0.04324449
   34  2  0 0.5  0  0 0.5 1 1 1    0.00336356    0.00247767
   35  4  0 0.5  2  2 0.5 1 1 1    0.00004349    0.00563919
<< .... 25 lines deleted .... >>

   P(S) = 0.583931,  P(S') = 0.303015, P(P) = 0.045894, P(D) = 0.067025

  T:
   61  0  0 0.0  0  0 0.0 1 1 1    0.07138963    0.58052013
   62  2  0 0.0  1  1 0.0 1 1 1    0.00000000    0.00000000
   63  2  0 0.0  0  0 0.0 1 1 1    0.02395303    0.29979838
   64  2  1 1.0  1  1 0.0 1 1 1    0.00005476    0.04324449
   65  4  0 0.0  2  2 0.0 1 1 1    0.00000006    0.00202213
<< .... 25 lines deleted .... >>

   P(S) = 0.884924,  P(S') = 0.002022, P(P) = 0.045894, P(D) = 0.067137

   norm of X channels =  0.999866
   norm of Y channels =  0.999866
   norm of T channels =  0.999977

  JJ coupling: X
  [ 0 1/2 + 0 0/2 ]0.5,  0.5;  0.0 :    0.58393105
  [ 1 1/2 + 1 2/2 ]0.5,  0.5;  0.0 :    0.23452232
  [ 1 3/2 + 1 2/2 ]0.5,  0.5;  0.0 :    0.10610143
  [ 1 1/2 + 1 2/2 ]0.5,  0.5;  0.0 :    0.02629508
  [ 1 3/2 + 1 2/2 ]0.5,  0.5;  0.0 :    0.03494796
  Prob(10be in state 1) =  0.930569 from Jab  0.000000  0.930569  0.000000
  Prob(10be in state 2) =  0.069297 from Jab  0.000000  0.069297  0.000000
  Total norm in jj basis =  0.999866

  JJ coupling: Y
  [ 0 0/2 + 0 1/2 ]0.5,  0.5;  0.0 :    0.58393105
  [ 1 2/2 + 1 1/2 ]0.5,  0.5;  0.0 :    0.02139192
  [ 1 2/2 + 1 3/2 ]0.5,  0.5;  0.0 :    0.31923183
  [ 1 2/2 + 1 1/2 ]0.5,  0.5;  0.0 :    0.01520326
  [ 1 2/2 + 1 3/2 ]0.5,  0.5;  0.0 :    0.04597204
  Prob(10be in state 1) =  0.930569 from Jab  0.000000  0.930569  0.000000
  Prob(10be in state 2) =  0.069297 from Jab  0.000000  0.069297  0.000000
  Total norm in jj basis =  0.999866

  JJ coupling: T
  [ 0 1/2 + 0 1/2 ]0.0,  0.0;  0.0 :    0.88492424
  [ 1 1/2 + 1 1/2 ]0.0,  0.0;  0.0 :    0.02908190
  [ 1 3/2 + 1 3/2 ]0.0,  0.0;  0.0 :    0.01454095
  [ 0 1/2 + 2 3/2 ]2.0,  2.0;  0.0 :    0.02071231
  [ 0 1/2 + 2 5/2 ]2.0,  2.0;  0.0 :    0.03106845
  Prob(10be in state 1) =  0.930569 from Jab  0.930569  0.000000  0.000000
  Prob(10be in state 2) =  0.069408 from Jab  0.000000  0.000000  0.069408
  Total norm in jj basis =  0.999977

 Ground state  0.0+ ENGY =   -4.877592
\end{verbatim}
\normalsize
\subsection{$^6$He}
Input file {\em he6psk06.in} and complete output file {\em
he6psk06.out} are provided in the electronic file distribution.
This models two neutrons outside an inert $^4$He core with a
Gaussian-shape n-$^4$He potential, with SUSY elimination of the
0s bound state in the s-wave potential, and a GPT nn potential,
as in \cite{dan98,thom00}.

\subsection{$^8$B}
Input file {\em b8grig1k6.in} and complete output file {\em
b8grig1k6.out} are provided in the electronic file distribution.
This models $^8$B as a three-body cluster of $^3$He, $^4$He and
a proton, as in \cite{grig}.

\begin{acknowledgments}

This work was supported  by NSCL-Michigan State University
(U.S.A.), FCT grant POCTI/36282/99 (Portugal),  EPSRC grant
GR/M/82141 (U.K.) and Russian grant RFBR 02-02-16174.

\end{acknowledgments}


\begin{thebibliography}{99}
\bibitem{be11} F.M. Nunes, I.J. Thompson and R.C. Johnson, Nucl. Phys. A 596 (1996) 171.
\bibitem{c19} D. Ridikas et al., Nucl. Phys. A 628 (1998) 363.
\bibitem{phyrep} M. V. Zhukov, B. V. Danilin, D. V. Fedorov, J. M. Bang,
I. J. Thompson and J. S. Vaagen, Phys. Rep. {\bf 231} (1993) 151.
\bibitem{li11} I.J. Thompson and M.V. Zhukov, Phys. Rev. C 49 (1994) 1904.
\bibitem{he8} M. Meister {\it et al.}, Nucl. Phys. A 700 (2002) 3.
\bibitem{be12} F.M. Nunes, J.A. Christley, I.J. Thompson, R.C. Johnson, V.D. Efros,
Nucl. Phys. A 609 (1996) 43.
\bibitem{reactions} J.S. Al-Khalili and F.M. Nunes, Topical Review J. Phys. G.  29 (2003) R89.
\bibitem{jim} J.S. Al-Khalili  and J.A. Tostevin, Phys. Rev. Lett. 76 (1996) 3903.
\bibitem{ganil} S. Fortier {\it et al.}, Phys. Lett. B 461 (2001) 99;
    J. Winfield {\it et al.}, Nucl. Phys. A 683 (2001) 48.
\bibitem{crespo} R. Crespo and I.J. Thompson, Nucl. Phys. A 689 (2001) 559;
    R. Crespo, I.J. Thompson and A. Korsheninnikov, Phys. Rev. C 66 (2002) 021002R.
\bibitem{faddeev} L.D. Faddeev, JETP 39 (1960) 1459.
\bibitem{gronwall} T.H. Gronwall, Phys. Rev. {\bf 51} (1937) 655.
\bibitem{delves} L.M. Delves, Nucl. Phys. {\bf 20} (1962) 268.
\bibitem{ray70}
J. Raynal and J. Revai, Nuovo Cimento {\bf A68} (1970) 612.
\bibitem{brink} D.M. Brink, G.R. Satchler, {\em Angular Momentum},
Clarendon (Oxford) 1994.
\bibitem{dan98} B.V. Danilin, I.J. Thompson, M.V. Zhukov and J.S. Vaagen,
Nucl. Phys. {\bf A632} (1998) 383.
\bibitem{thom00} I.J. Thompson, B.V. Danilin, V.D. Efros, J.S. Vaagen,
J.M. Bang and M.V. Zhukov, Phys. Rev. C {\bf 61} (2000) 24318.
\bibitem{ss}
J.-M. Sparenberg and D. Baye, Phys. Rev. Lett. {\bf 79} (1997) 3802.
\bibitem{feshbach} H. Feshbach, Ann. Phys. {\bf 5} (1958) 357.
\bibitem{sturmxx} I.J. Thompson (2002), Program {\tt sturmxx}, Users
manual available from the author.
\bibitem{grig}
L.V. Grigorenko, B.V. Danilin, V.D. Efros, N.B. Shul'gina,
M.V. Zhukov, Phys. Rev. C {\bf 57} (1998) R2099; {\bf 60}
(1999) 044312
\end{thebibliography}
\end{document}